\documentclass[epj,dvips]{webofc}
\usepackage[varg]{txfonts}   
\providecommand{\tabularnewline}{\\}
\newcommand{\aap}{A\&A}

\newcommand{\apj}{ApJ}
\newcommand{\apjl}{\apj}

\newcommand{\apjs}{ApJS}

\woctitle{Physics at the Magnetospheric Boundary}
\begin{document}
\title{Oscillations of the Boundary Layer and High-frequency QPOs}
%
%

\author{A.~A. Blinova\inst{1}\fnsep\thanks{\email{alisablinova@gmail.com}}, M. Bachetti\inst{2},
M.~M. Romanova\inst{1}}

\institute{Department of Astronomy, Cornell University, Ithaca, NY
14853-6801, USA \and Institut de Recherche en Astrophysique et
Planetologie, Toulouse, 31400, France}

\abstract{%
We observed persistent high-frequency oscillations of the boundary
layer near an accreting, weakly-magnetized star in global 3D MHD
simulations. The tilted dipole magnetic field is not strong enough
to open a gap between the star and the disk. Instead, it forms a
highly-wrapped azimuthal field near the surface of the star which
slows down rotation of the disk matter, while a small tilt of the
field excites oscillations of the boundary layer with a frequency
below the Keplerian frequency. This mechanism may be responsible
for the high-frequency oscillations in accreting neutron stars,
white dwarfs and classical T Tauri stars.}
\maketitle
\vspace{-0.1cm}\section{Introduction} \label{intro}

Accretion through a boundary layer is expected in the cases where
a star has a relatively weak magnetic field, so that the accretion
disk reaches the surface of the star. In most investigations of
the boundary layer the magnetic field has been neglected
(e.g.~\cite{popham1995,fisker2005,inogamov1999,piro2005,paczynski1978}).

It is important to study the boundary layer in situations where a
weak magnetic field is present. Paczy\'nski suggested that the
magnetic field may influence the dynamics of the boundary layer,
forming a relatively thick torus \cite{paczynski1978}. He
suggested that the high-frequency oscillations in the Dwarf
Novae-type white dwarfs may be connected with oscillations of the
weakly-magnetized, rotating torus \cite{paczynski1978}. The weak
magnetic field may also influence boundary layer accretion in
young stars, and determine the highest-frequency QPOs in accreting
neutron stars in Low-mass X-ray Binaries (LMXBs)
\cite{vanderklis2006}.





Global 3D MHD simulations of accretion to a star with a weak
magnetic field have shown that even a small magnetosphere
influences the dynamics of the inner disk \cite{romanova2009}.
Accretion in the boundary layer regime has also been studied in
the cases where accretion in the disk is driven by the
magnetorotational (MRI) instability
\cite{romanova2012,balbus1991}. Simulations have shown that a
star's weak magnetic field is enhanced due to the rotation of
matter in the inner disk, and this field influences boundary layer
accretion.

The goal of our research was to perform long-lasting 3D MHD
simulations of the boundary layer in weakly-magnetized stars, and
to investigate possible high-frequency variabilities connected
with this type of accretion. In our simulations we discovered
high-frequency, quasi-periodic oscillations (QPOs) of the boundary
layer. Such oscillations are expected in different types of stars.
Here, we show an example of boundary layer oscillations in
accreting neutron stars.

\vspace{-0.1cm}\section{Model} \label{method}

To investigate the boundary layer, we performed multiple 3D MHD
(magnetohydrodynamic) simulations using a Godunov-type
(e.g.~\cite{powell1999}) numerical code, written in a  ``cubed
sphere" coordinate system rotating with the star
\cite{koldoba2002}. We solved 8 partial differential equations for
density, entropy, three components of momentum and three
components of the magnetic field. Simulations were performed using
a high grid resolution (244 grid points in the azimuthal
direction). Initial and boundary conditions are described, e.g.,
in \cite{romanova2012}.

We took the mass of a simulated neutron star to be
$M_\star=1.4M_{\odot}$, dipole magnetic field $B=2\times10^7 G$,
and radius of the star $R_\star=10$km. We considered accretion
from an $\alpha$-disk \cite{shakura1973} with $\alpha$-parameter
$\alpha=0.04$. We used the Nowak-Wagoner potential
\cite{nowak1991} to approximate the general relativistic effects:
$$\Phi_{\rm NW}=-\frac{G M_\star}{r}\bigg[1-3\frac{G M_\star}{r
c^2}+12\bigg(\frac{G M_\star}{r c^2}\bigg)^2\bigg]=-\frac{G
M_\star}{r}\bigg[1-\frac{3r_g}{2r}+3\bigg(\frac{r_g}{r}\bigg)^2\bigg]~,$$
where $r_g={2GM_\star}/{c^2}=4.15$km is the Schwarzschild radius.
This potential gives a reasonable approximation of general
relativistic effects near the star (better, e.g., than the
Paczy\'nski-Wiita potential ~\cite{abramowicz2009}).




\vspace{-0.1cm}\section{Results}

\begin{figure}[!ht]
\centering
\includegraphics[width=5cm,clip]{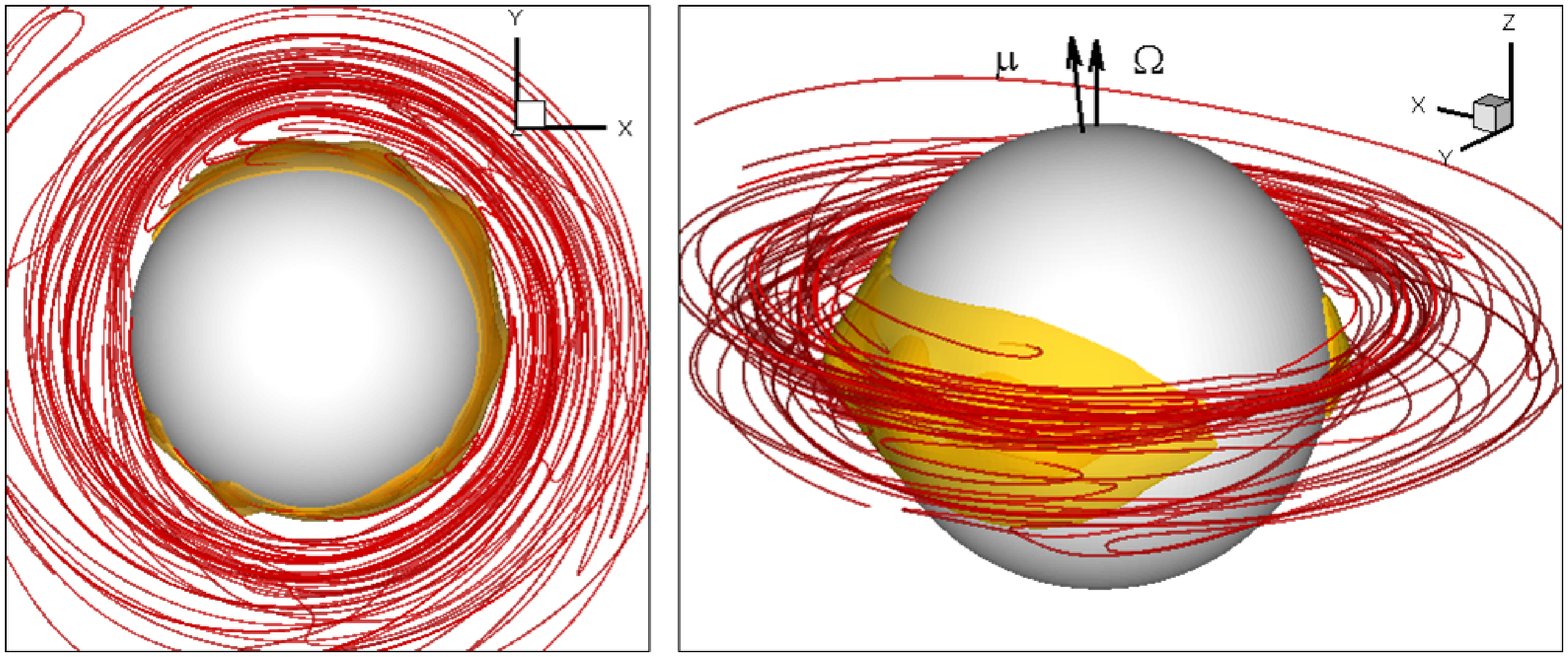}
\includegraphics[width=9cm,clip]{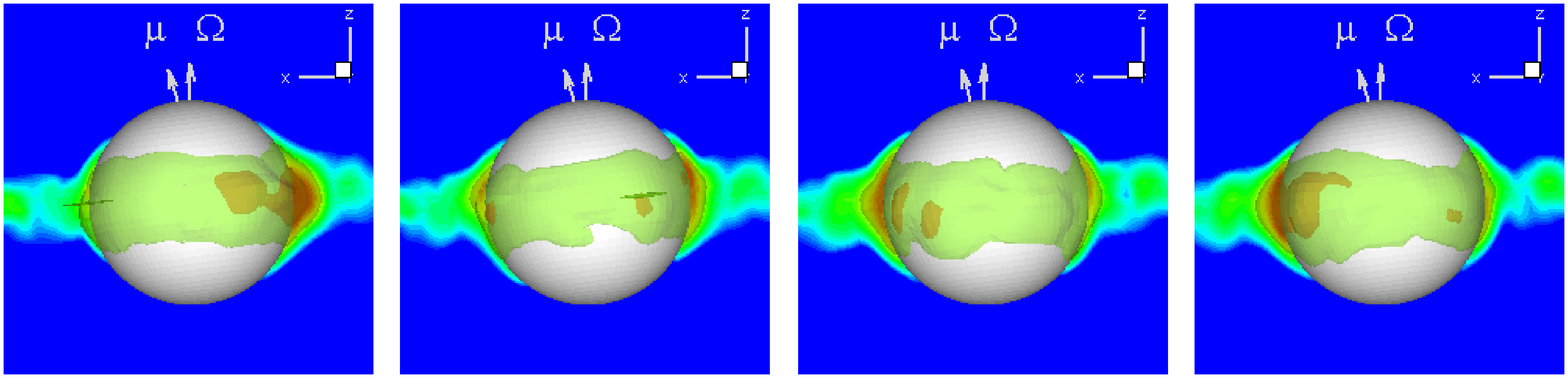}
\vspace{-0.2cm}\caption{\textit{Left two panels}: highly-wrapped
azimuthal field lines in the inner disk. \textit{Right four
panels}: 3D side views of boundary layer oscillations, shown over
four consecutive moments in time. Images are shown in a coordinate
system rotating with the star.}
\label{3d-side-main-800-806}       
\end{figure}

From our simulations we observed a boundary layer which is
relatively thick, ($0.05-0.1)R_\star$, due to the interaction of
the inner disk matter with the weak magnetic field. Our
simulations show persistent, high-frequency quasi-periodic
oscillations (QPOs) of matter in the boundary layer.
Fig.~\ref{3d-side-main-800-806} (right four panels) shows an
example of variation in the density distribution of the boundary
layer over four consecutive moments in time. A density enhancement
is formed and rotates with sub-Keplerian frequency. 
Highly-wrapped azimuthal field lines are formed in the inner disk,
and are responsible for the relatively high thickness of the
boundary layer (see Fig.~\ref{3d-side-main-800-806}, left two
panels).

\begin{figure}[!ht]
\centering
\includegraphics[width=6cm,clip]{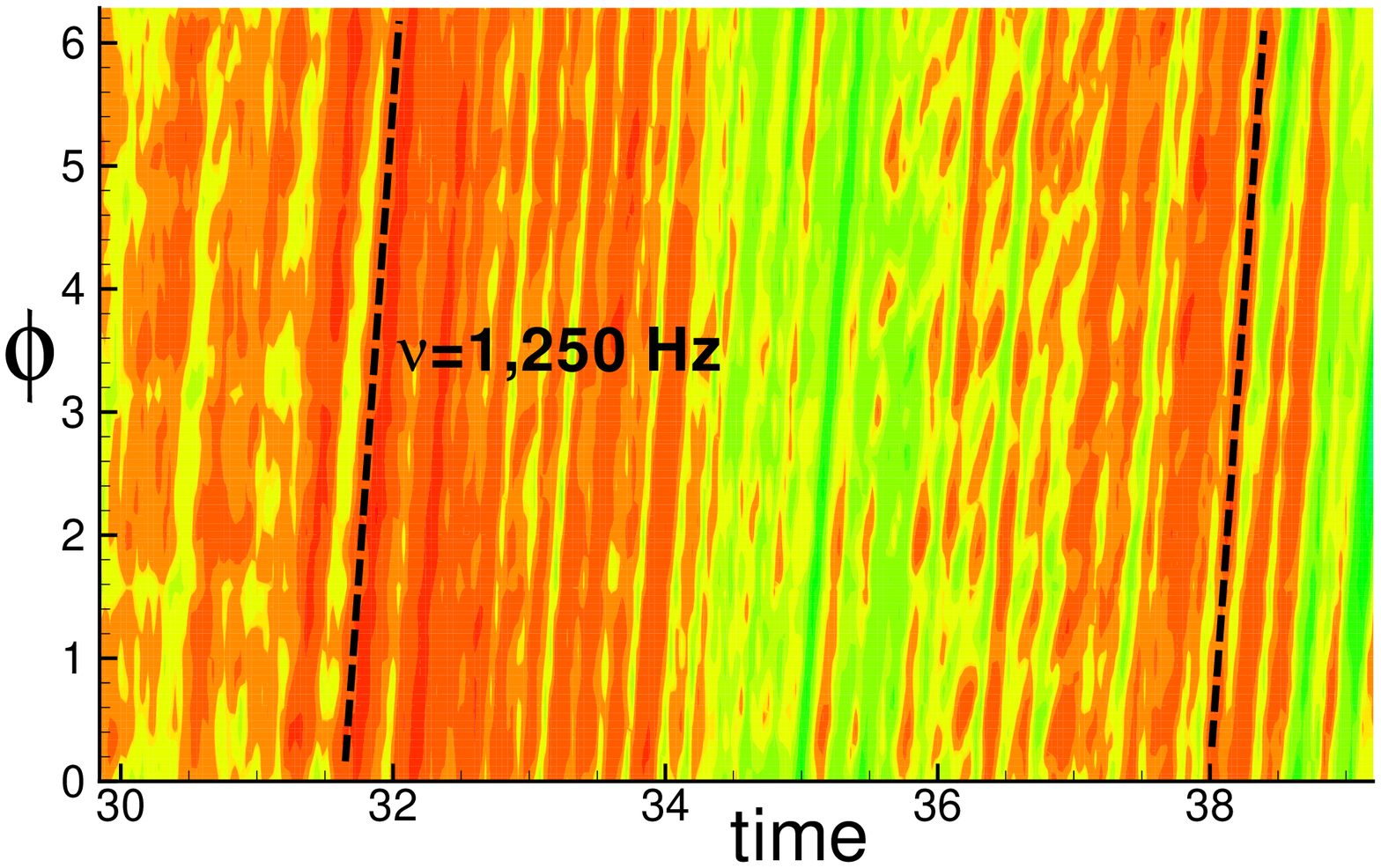}
\includegraphics[width=7cm,clip]{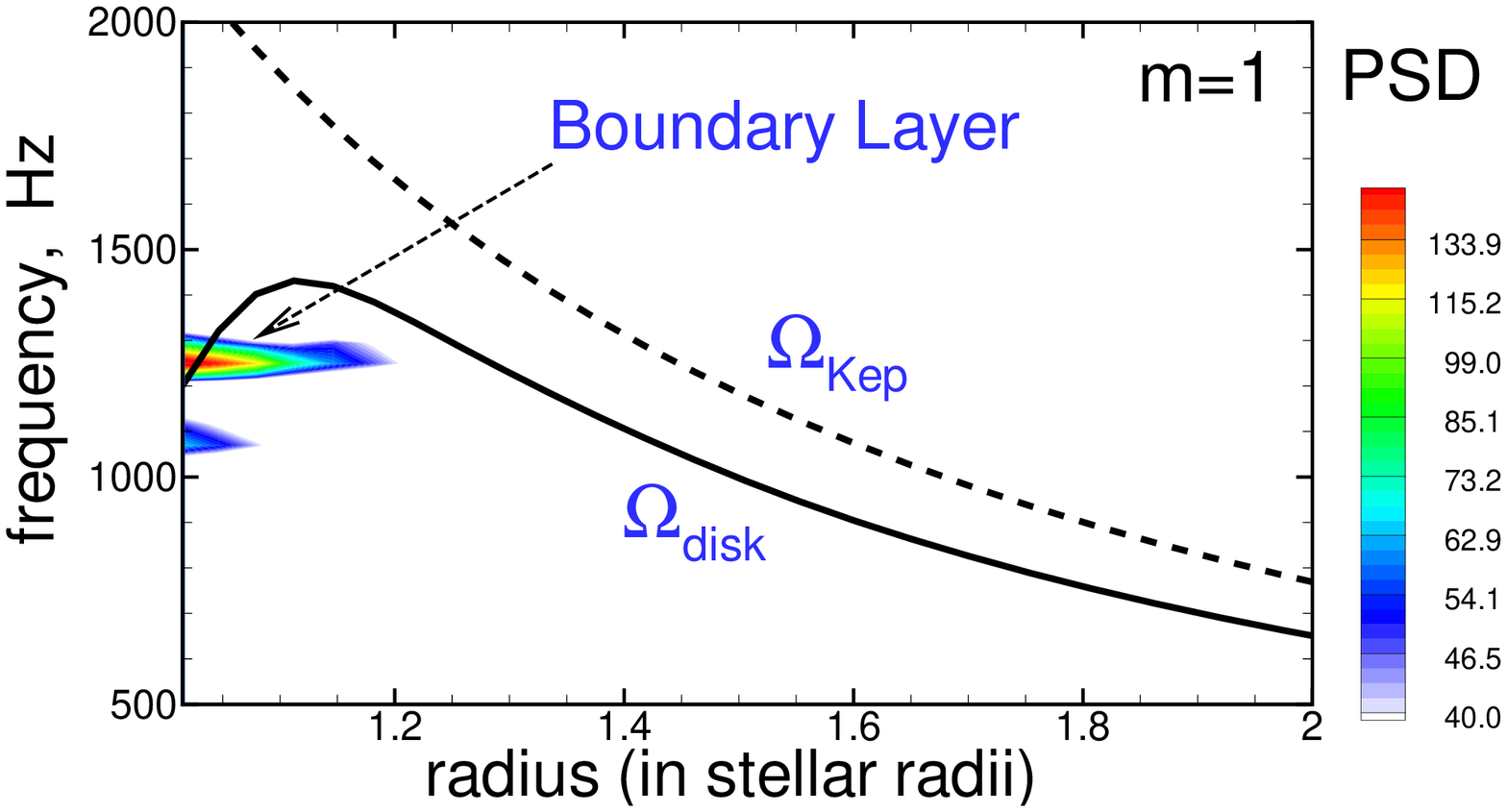}
\vspace{-0.2cm}\caption{\textit{Left panel}: a spot-omega diagram
\cite{kulkarni2008} of density distribution in the equatorial
plane over $\phi$ (azimuthal direction) at the radius of
$r=1.1R_\star$ for different moments in time. Time is given in the
units of one orbital period. The slope of the lines represents the
frequency of boundary layer oscillations. \textit{Right panel}:
Power Spectral Density (PSD) is calculated for the surface density
distribution near the star. The m=1 (one-armed) density wave
dominates and has a frequency of $\nu\approx1,250$Hz.}
\label{spot-omega-main-5}       
\end{figure}

To find the frequency of the oscillations, we performed different
analyses. Fig.~\ref{spot-omega-main-5} (left panel) shows a
spot-omega diagram (see, e.g., \cite{kulkarni2008}) of density
distribution in the boundary layer, where the slope of the lines
represents the frequency of the QPOs: $\nu\approx1,250$Hz. The
Power Spectral Density analysis has shown the presence of a
one-armed (m=1) density wave rotating at the same frequency (see
Fig.~\ref{spot-omega-main-5}, right panel). This frequency is
lower than the Keplerian frequency at the surface of the star
($\nu_{\rm Kep}\approx2,200$Hz), but is much higher than the
frequency of the star ($\nu_\star=186$Hz). This frequency is
located below the maximum in the $\Omega_{\rm disk}$ distribution,
and therefore the observed wave may represent a radially-trapped
density wave \cite{lovelace2009}.

The frequency of $\nu\approx1,250$Hz obtained in our simulations
should be corrected, taking into account the fact that we used the
Nowak-Wagoner potential instead of the exact GR approach. In the
case of a Keplerian disk (no magnetic field), the frequency
$\nu_{\rm NW}$ is different from the Keplerian frequency $\nu_{\rm
Kep}=({1}/{2\pi})({GM_\star}/{r^3})^{1/2},$ and is: $\nu_{\rm
NW}=\nu_{\rm Kep}(1-3\delta+9\delta^2)^{1/2},$ where
$\delta=r_g/r$. However, in Schwarzschild geometry, the frequency
$\nu_{S}=\nu_{\rm Kep}$ exactly. Taking into account the
correction $\nu_{S}=\nu_{\rm NW}(1-3\delta+9\delta^2)^{-1/2}$ at a
typical boundary layer radius of $r=1.1R_\star$, we obtain
$\nu_{S}=1,170$Hz. On the other hand, a star has a spin
$\nu_\star=186$Hz, and a small correction for the Kerr geometry
will slightly increase the frequency of boundary layer
oscillations \cite{miller1998}.


\begin{figure} [!ht]
\centering
\includegraphics[width=6cm,clip]{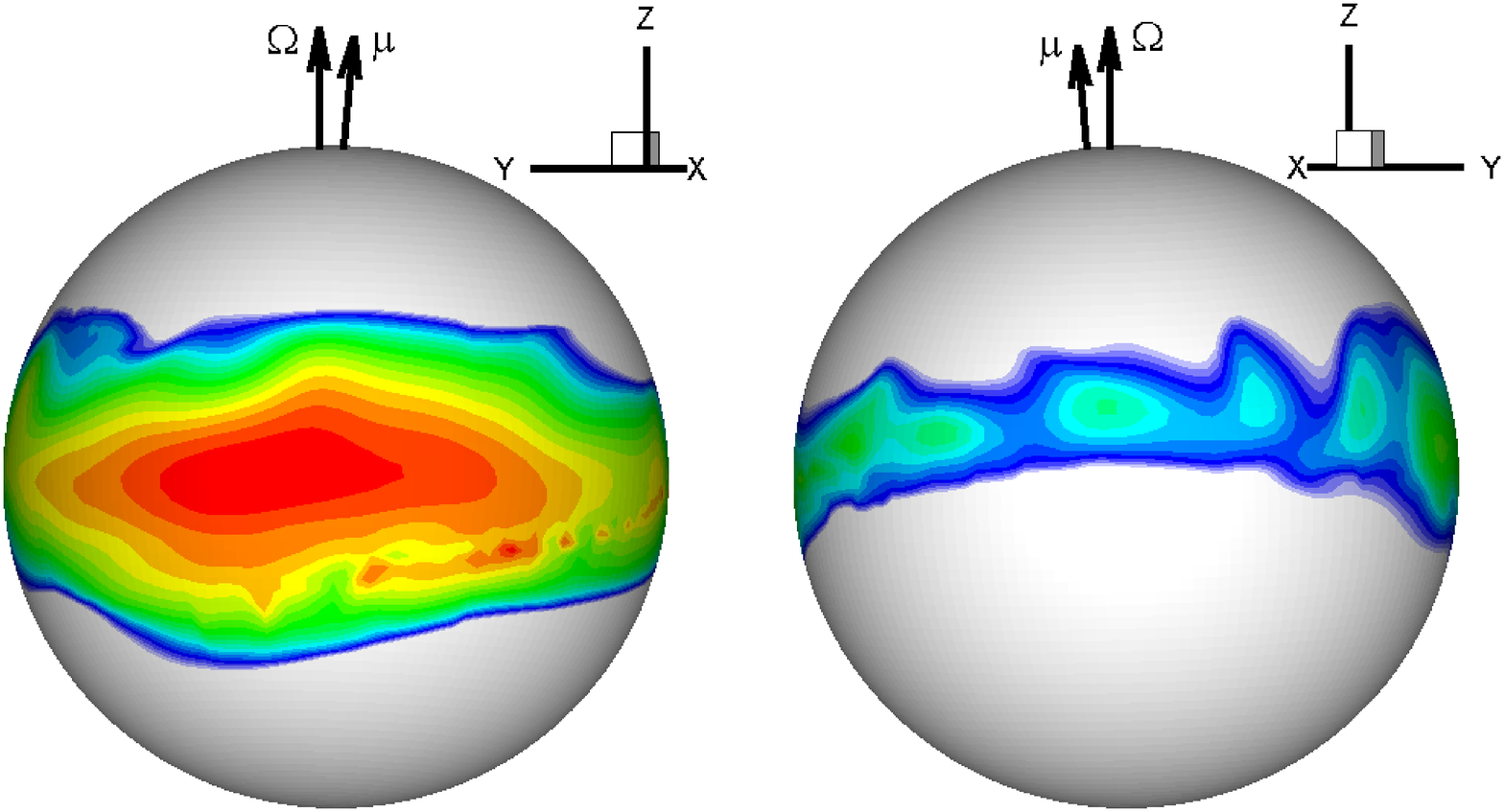}\hspace{0.5cm}
\includegraphics[width=6cm,clip]{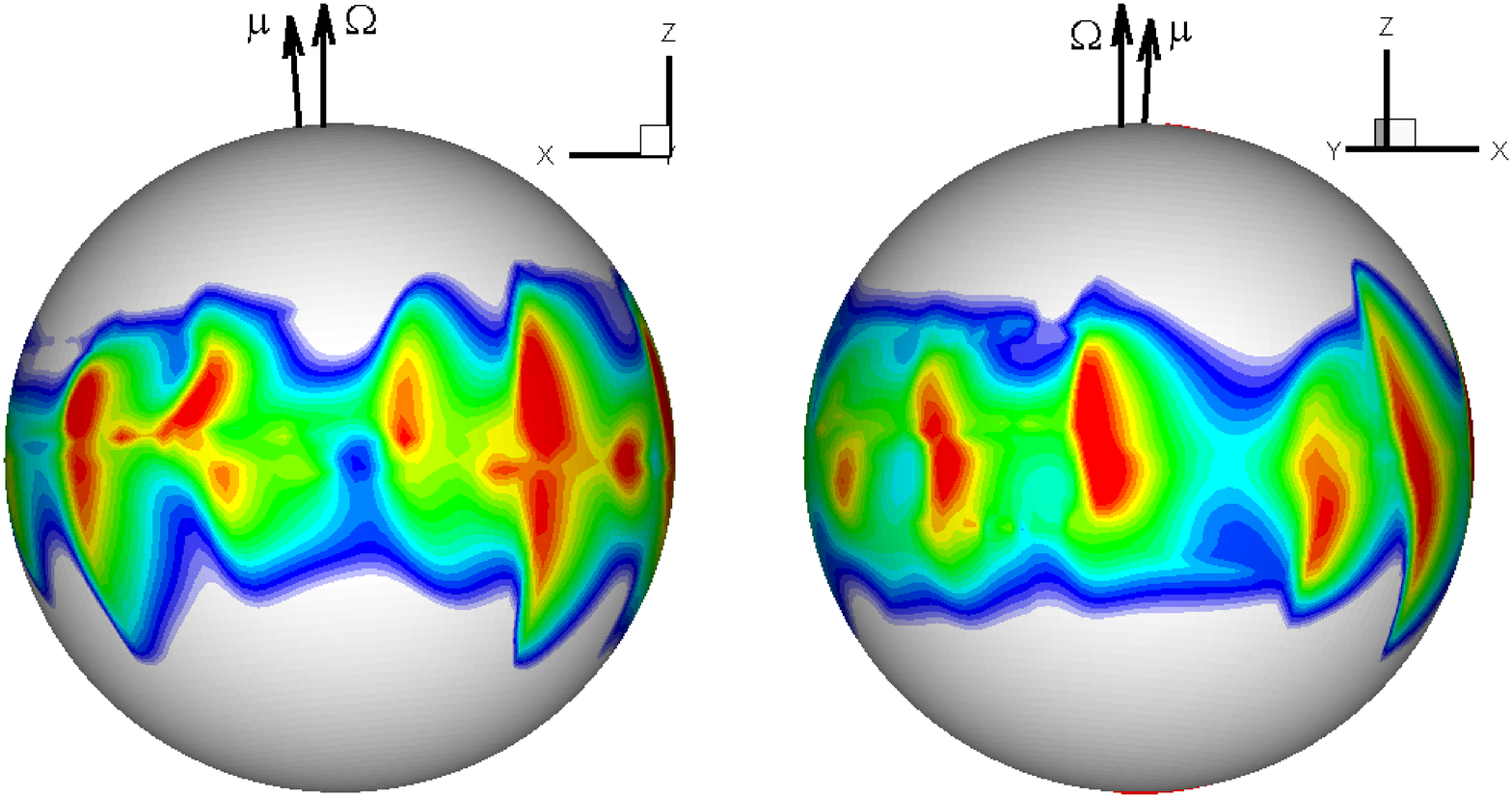}
\vspace{-0.2cm}\caption{\textit{Left two panels}: the dominant
spot of the boundary layer at time t=40, where time is measured in
periods of Keplerian rotation at $r=1\approx3R_\star$. The star is
shown at two different angles. \textit{Right two panels}: Two
examples (at two different moments in time) of smaller spots in
the boundary layer, which may be connected with the
Kelvin-Helmholtz instability.}
\label{spot-main-1000}       
\end{figure}

The boundary layer is adjacent to the star, however, the density
wave in the boundary layer rotates much more rapidly than the
star. The density distribution in the boundary layer right above
the surface of the star reflects the propagation of this density
wave. We observe one large spot which rotates with the frequency
of the boundary layer (see Fig.~\ref{spot-main-1000}, left two
panels). Also, we often see several smaller spots in the boundary
layer (see Fig.~\ref{spot-main-1000}, right two panels). These
spots are probably connected with the Kelvin-Helmholtz
instability, which could arise due to the velocity shear between
the rapidly-rotating boundary layer and the slower-rotating layers
above the surface of the star. We suggest that the observed
boundary layer oscillations may be excited by the small tilt
($\Theta=5^\circ$) of the magnetosphere. Such a tilt is expected
in many realistic systems.

The highest frequencies of QPOs observed in LMXBs are
$\sim1,000-1,250$Hz \cite{vanderklis2006}. In some models, it is
suggested that these frequencies are associated with Keplerian
frequency $\nu_{\rm
ISCO}=1,564(M_\star/1.4M_\odot)^{1/2}(R_\star/10\rm km)^{-3/2}\rm
Hz$ at the last stable orbit $r_{\rm
ISCO}=3r_g=12.45(M_\star/1.4M_\odot)\rm km,$ which gives a
possible way of deriving the mass to radius ratio for neutron
stars \cite{boutelier2009,barret2011}. In our research, we show
that in the presence of a weak magnetic field the rotation near a
star is sub-Keplerian, and that the highest-frequency QPOs may
instead be connected with the oscillations in the boundary layer.


\vspace{-0.1cm}\section{Summary} \label{summary}

We ran 3D MHD simulations of accretion to a weakly-magnetized
neutron star of mass $M_\star=1.4M_\odot$ and radius $r=10$km,
accreting in the boundary layer regime. We found persistent
high-frequency quasi-periodic oscillations of the boundary layer
which are associated with the propagation of a one-armed density
wave. The frequency of these oscillations, $\nu\approx1,170$Hz, is
lower than the Keplerian frequency at $R_\star$ and at $r_{\rm
ISCO}$. These oscillations should be systematically studied for
different parameters of stars and their magnetic fields.

\section*{Acknowledgment}
We thank the conference organizers, especially the chair, Dr. E.
Bozzo, for the excellent meeting. Resources supporting this work
were provided by the NASA High-End Computing (HEC) Program through
the NASA Advanced Supercomputing (NAS) Division at Ames Research
Center and the NASA Center for Computational Sciences (NCCS) at
Goddard Space Flight Center. The research was supported by NASA
grant NNX11AF33G and NSF grant AST-1008636.

%
%


\end{document}